\documentclass[sigconf]{acmart}
\usepackage{xcolor}
\usepackage{listings}
\usepackage[scaled]{beramono}
\usepackage[T1]{fontenc}
\renewcommand\footnotetextcopyrightpermission[1]{} 

\newcommand{\tabincell}[2]{\begin{tabular}{@{}#1@{}}#2\end{tabular}}  

\AtBeginDocument{%
  \providecommand\BibTeX{{%
    \normalfont B\kern-0.5em{\scshape i\kern-0.25em b}\kern-0.8em\TeX}}}

\begin{document}
\pagestyle{plain}
\pagenumbering{gobble}
\title{HybridDNN: A Framework for High-Performance Hybrid DNN Accelerator Design and Implementation}


\author{ 
Hanchen Ye$^{1}$, Xiaofan Zhang$^{1}$, Zhize Huang$^{2}$, Gengsheng Chen$^{2}$, Deming Chen$^{1}$
}
\affiliation{ 
	\institution{$^1$University of Illinois at Urbana-Champaign, $^2$Fudan University}
}
\affiliation{ \normalsize
	\textit{\{hanchen8, xiaofan3, dchen\}@illinois.edu, \{18212020085, gschen\}@fudan.edu.cn}
}







\begin{abstract}
\vspace{-2pt}
To speedup Deep Neural Networks (DNN) accelerator design and enable effective implementation, we propose HybridDNN, a framework for building high-performance hybrid DNN accelerators and delivering FPGA-based hardware implementations. Novel techniques include a highly flexible and scalable architecture with a hybrid Spatial/Winograd convolution (CONV) Processing Engine (PE), a comprehensive design space exploration tool, and a complete design flow to fully support accelerator design and implementation. 
Experimental results show that the accelerators generated by HybridDNN can deliver 3375.7 and 83.3 GOPS on a high-end FPGA (VU9P) and an embedded FPGA (PYNQ-Z1), respectively, which achieve a 1.8x higher performance improvement compared to the state-of-art accelerator designs. This demonstrates that HybridDNN is flexible and scalable and can target both cloud and embedded hardware platforms with vastly different resource constraints. 

\end{abstract}

\maketitle

\vspace{-6pt}
\section{Introduction}
\vspace{-2pt}
With deeper and more complicated layer connections, DNNs are becoming more compute- and memory-intensive, which require hardware accelerators to deliver high throughput, low end-to-end latency, and high energy efficiency. Recently, researchers have focused on building customized DNN accelerators by taking advantage of different hardware devices, including GPUs, FPGAs, and ASICs, to improve the speed and efficiency of DNN inference \cite{diannao,qiu2016going,lrcn,improving,dnnbuilder,clouddnn,zhang2020skynet,xu2020autodnnchip}. 
By considering the accelerator deployment for real-life AI applications, energy-hungry GPU-based accelerators are difficult to meet the energy/power constraints while the ASIC-based designs require a long time-to-market period. FPGAs, therefore, become promising candidates for DNN implementations with improved latency and energy consumption compared to GPUs while offering much more flexibility than ASICs because of their reconfigurable features \cite{qiu2016going,lrcn,improving,dnnbuilder,clouddnn}. 

Recent years, High-Level Synthesis (HLS) techniques have significantly improved the developing efficiency of FPGA-based hardware design by allowing to program FPGAs in high-level languages (e.g., C/C++)\cite{chen2009lopass,chen2005xpilot,rupnow2011high}.
However, building a high-performance FPGA-based DNN accelerator is still non-trivial since it requires
customized hardware implementation, iterative hardware/software verification to ensure functional correctness, and effective design space exploration for sophisticated accelerator configurations.
To improve the efficiency of accelerator design, we have witnessed a growing interest in developing automation frameworks for building DNN accelerators from a higher level of abstraction, using DNN-specific algorithmic descriptions and pre-defined high-quality hardware templates for fast design and prototyping \cite{wang2018design,dnnbuilder,clouddnn,deepburning,caffeine,fpdnn}. 
However, design difficulties still exist as recent development trends in cloud and embedded FPGAs present completely different challenges for satisfying diverse requirements of DNN applications. For example, latest-generation cloud FPGAs have widely utilized multiple dies to multiply the available resources for delivering higher throughput \cite{intel_emib,xilinx_ultrascale}. However, the cross-die routing and distributed on-chip memory can easily cause timing violations, and lower the achievable performance once the accelerator designs fail to scale up/down to match the die size. On the other hand, embedded FPGAs are integrating heterogeneous components (e.g., CPUs, GPUs) to handle different parts of the targeted tasks efficiently. Without a highly flexible task partitioning strategy, it is impossible to fully utilize the on-chip resources and leverage all the advantages of the particular hardware. 
%
Meanwhile, many researchers are seeking for improvements from a software perspective by using fast CONV algorithms (e.g., Winograd and Fast Fourier Transform)  \cite{winograd,evaluating,caffeinated,zhuge}. Although these accelerators can achieve higher performance than conventional designs, they suffer from more stringent restrictions imposed by use cases and require more complicated design flows. 

To address these challenges, we propose HybridDNN, an end-to-end framework of building and implementing high-performance DNN accelerators on FPGAs leveraging the Winograd algorithm. To summarize, the main contributions of this paper are as follows.
\leftmargini=4mm
\begin{itemize}
    \item \vspace{-2pt} 
    We propose HybridDNN, which can generate highly optimized accelerators for the latest generation of cloud and embedded FPGAs. Described by HLS, the generated designs can be easily fine-tuned for better customization.
    \item We introduce a highly flexible and scalable DNN accelerator architecture with a
    hybrid-mode (Spatial and Winograd) CONV PE and a multi-dataflow (with Input Stationary (IS) and Weight Stationary (WS)) structure.
    \item We integrate a comprehensive set of tools in HybridDNN for performance estimation and design space exploration to guide the accelerator design with improved performance.
\end{itemize}

\vspace{-8pt}
\section{Related Works}
\vspace{-2pt}
\label{sec:background}
There are intensive studies of designing and optimizing DNN accelerators on FPGAs. Authors in \cite{qiu2016going} present a dynamic data quantization scheme for DNN parameters to  relax the required memory access bandwidth.
To reduce DNN inference latency for both embedded and cloud platforms, DNNBuilder proposes a fine-grained, layer-based pipeline architecture along with optimal resource allocation targeting real-life DNN with high definition inputs \cite{dnnbuilder}. 
More advanced optimizations are investigated in accelerator design to achieve better balance between DNN inference accuracy and efficiency, such as using extremely low precision DNN parameters \cite{wang2018design}, fast CONV algorithms \cite{zhuge}, and hardware/software co-design \cite{jiang2019accuracy,hao2019fpga,zhang2020skynet}. The literature also focuses on developing systematic tools for building DNN accelerators. A framework is proposed in \cite{wei2017automated} to use systolic arrays for accelerating DNN inference and the framework proposed in \cite{zeng2018framework} enables task partitioning with compute-intensive CONV layers implemented on an FPGA and fully-connected (FC) layers handled by a CPU. In addition, DeepBurning \cite{deepburning} introduces a design flow of using parameterized pre-defined RTL modules to construct accelerators and FP-DNN \cite{fpdnn} uses a mixture of RTL and HLS for better flexibility.

\vspace{-6pt}
\section{The Proposed Design Flow}
\begin{figure}[t!]
    \centering
    \vspace{-20pt}
    \includegraphics[width=0.48\textwidth]{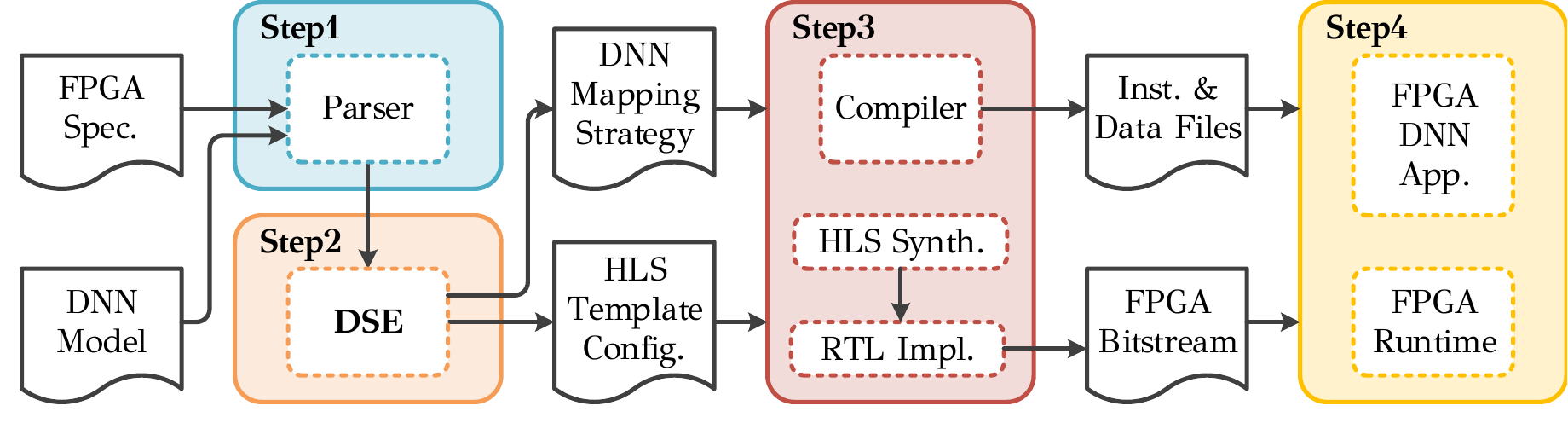}
    \vspace{-20pt}
    \caption{HybridDNN Design Flow}
    \label{fig:tool_chain}
    \vspace{-15pt}
\end{figure}

\vspace{-2pt}
HybridDNN provides an end-to-end design framework which can generate high-performance instruction-based DNN accelerator designs and FPGA implementations in four steps (Figure \ref{fig:tool_chain}). In \textbf{Step (1)}, the targeted FPGA specification and the pretrained DNN model are passed to HybridDNN parser to capture hardware resource availability and DNN structure.
In \textbf{Step (2)}, HybridDNN launches the Design Space Exploration (DSE) engine for design guidelines by considering both hardware and software perspectives. Design choices related to configurations of hardware instances on FPGAs are considered as hardware perspectives (e.g., the parallel factors of PE, the tile size of Winograd CONV) 
while factors corresponding to runtime task mapping are considered as software perspectives (e.g., selections of PE operating mode, dataflow strategies, and task partitioning schemes). Detailed discussions of the DSE will be in Section \ref{sec:dse}.
After running DSE, in \textbf{Step (3)}, the HLS template configurations are finalized and transformed into synthesizable C-level descriptions and ready for deployment on the targeted FPGA. The DNN mapping strategy is handled by HybridDNN compiler to generate executable instructions for running the generated accelerators which will be discussed in Section \ref{sec:accelerator}. 
In \textbf{Step (4)}, a light-weight runtime is deployed on the host CPU to manage the execution of the generated accelerator by enabling a 4-stage instruction pipeline and  I/O data management.

\vspace{-6pt}
\section{Accelerator Design}
\vspace{-2pt}
\label{sec:accelerator}
HybridDNN generates a hybrid Spatial/Winograd DNN accelerator and provides solutions to two major challenges of using such a hybrid design: (1) HybridDNN efficiently reuses one PE for both conventional CONV (Spatial CONV) and Winograd CONV to minimize the overhead of FPGA computing resources and (2) HybridDNN provides a decent input/output data organization mechanism to support a flexible switch between the Spatial and Winograd modes. We will illustrate why these challenges are non-trivial and our detailed solutions in this section.

\vspace{-6pt}
\subsection{Architecture and Instruction Design}
\begin{figure*}[t!]
\centering
    \vspace{-30pt}
    \includegraphics[width=\textwidth]{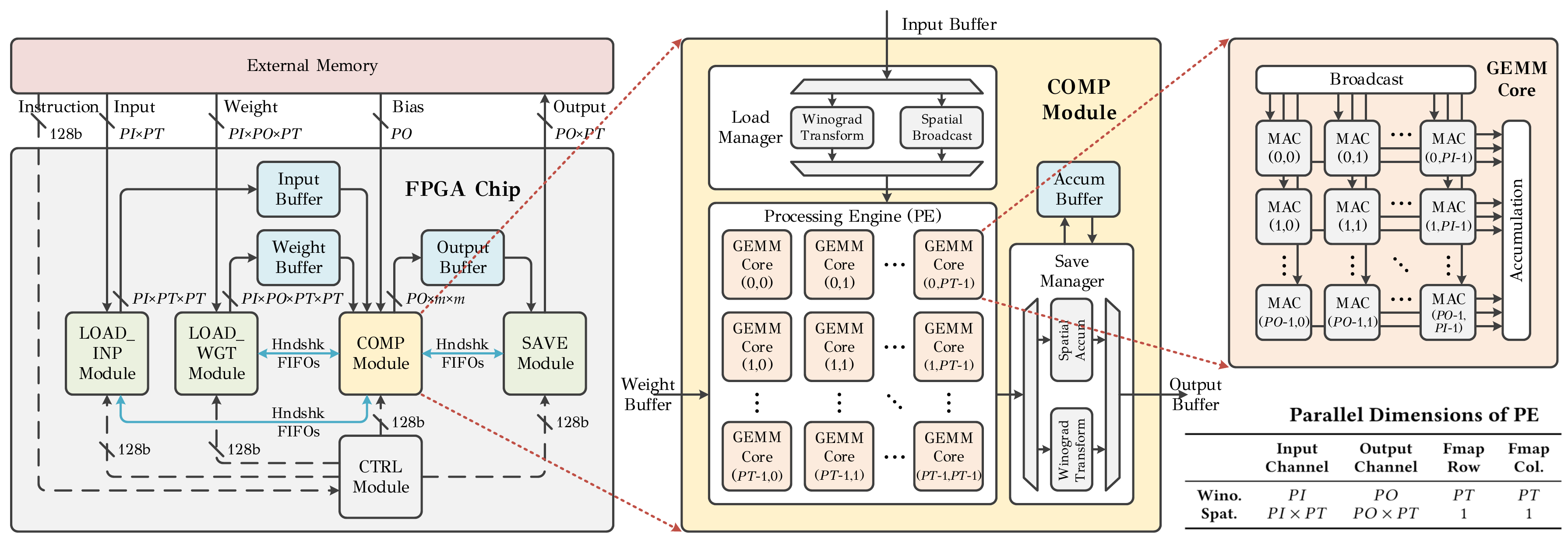}
    \vspace{-20pt}
    \caption{Hardware Architecture Design Generated by HybridDNN}
    \label{fig:system_arch} 
    \vspace{-11pt}
\end{figure*}

\vspace{-2pt}
\label{subsec:isa}
A folded accelerator architecture is generated by HybridDNN to maximize the support of different DNNs. 
As shown in Figure \ref{fig:system_arch}, the accelerator is constructed with four functional modules as: (1) a LOAD\_INP and (2) a LOAD\_WGT module for loading input feature maps and DNN parameters, respectively, from external memory to on-chip buffers; (3) a COMP module to handle computations; and (4) a SAVE module for sending intermediate results back to external memory. In addition, a controller (CTRL module) is designed for instruction fetch and decode. Solid lines and dash lines in Figure \ref{fig:system_arch} represent the data path and instructions path, respectively.

\begin{figure}
    \centering
    \vspace{-27pt}
    \includegraphics[width=0.48\textwidth]{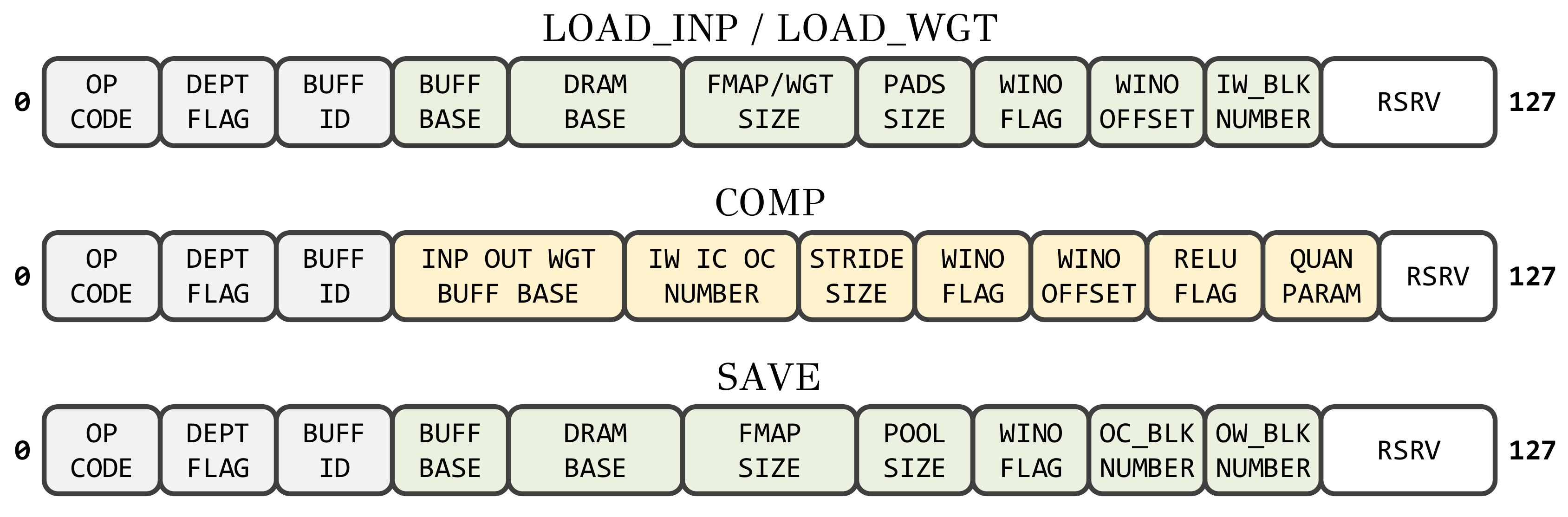}
    \vspace{-21pt}
    \caption{Customized Instruction Set}
    \vspace{-15pt}
    \label{fig:isa}
\end{figure}

To utilize these functional modules, we propose five different instructions as: LOAD\_INP, LOAD\_WGT, LOAD\_BIAS, COMP, and SAVE (Figure \ref{fig:isa}). Each instruction is encoded using 128 bits 
and contains a WINO\_FLAG domain for indicating the current CONV mode (whether Winograd or Spatial CONV). BUFF\_BASE and DRAM\_BASE domains in LOAD\_INP, LOAD\_WGT, and SAVE instructions allow the HybridDNN compiler to fully control the data access behavior of the accelerator and dynamically employ Input Stationary (IS) or Weight Stationary (WS) dataflow (with more details in Subsection \ref{subsubsec:dataflow}) following the exploration of DSE.

In order to maximize performance, we employ two approaches to let all the functional modules work concurrently. We first allocate ping-pong buffers for input/output data from/to the external memory to overlap data access and computation. Then, we introduce handshake FIFOs between three pairs of data producer and consumer, such as ``LOAD\_INP and COMP'', ``LOAD\_WGT and COMP'', and ``COMP and SAVE'' (which are indicated as blue arrows in Figure \ref{fig:system_arch}), to prevent hazards.
For each pair, the consumer will wait for the producer to emit a token through the handshake FIFO before reading and processing corresponding data. Meanwhile, the producer will wait for a token from the consumer as well, to avoid data pollution.
With these two approaches, we can effectively hide the external memory access latency and improve overall performance.

\vspace{-6pt}
\subsection{Hybrid Spatial and Winograd PE}
\vspace{-2pt}
\label{sec:engine}
The COMP module can dynamically reuse one PE to process either Spatial or Winograd CONV. As shown in Figure \ref{fig:system_arch}, the proposed COMP module allocates most of the computation resources to a PE with dynamic parallel factors according to different CONV modes. It also contains an accumulating buffer and a couple of load and save managers which can be reconfigured to satisfy different data access and computation patterns of Spatial or Winograd CONV.



\vspace{-4pt}
\subsubsection{\textbf{Winograd CONV}}
Assuming a convolutional layer with a 3-dim input feature $D$ (size $H \times W$ with $C$ channels) and a 4-dim kernel $G$ (size $R \times S$ with $K$ output and $C$ input channels), 
an $F(m \times m, r \times r)$ Winograd algorithm \cite{winograd} can generate output as Eq. \ref{equ:wino_tile}.
In this equation, $Y$ and $d$ represent output and input tiles, while $g$ represents kernels. $A$, $G$, and $B$ are constant transforming matrices and $\odot$ represents Element-Wise Matrix Multiplication (EWMM).
Input feature $D$ is partitioned into multiple input tiles, $d$, with size $(m+r-1) \times (m+r-1)$ while adjacent tiles share an $(r-1)$ overlap. In Eq. \ref{equ:wino_tile}, the output tile size is $m \times m$ and the kernel size is $r \times r$.
\vspace{-3pt}
\begin{equation}
\small
    Y=A^T \left[ \left[ GgG^T \right] \odot \left[ B^TdB \right] \right] A
    \label{equ:wino_tile}
    \vspace{-3pt}
\end{equation}
The advantage of Winograd CONV comes from the lower computation complexity. For example, an $F(4 \times 4, 3 \times 3)$ Winograd algorithm requires 36 multiplications for one output tile, while the Spatial CONV needs 144 multiplications. The reduction of multiplications is 4 times in this case. Although Winograd CONV introduces extra additions due to the transformation in Eq. \ref{equ:wino_tile}, the cost of implementing addition is much lower than multiplication in hardware.
These extra additions will not cause obvious performance slowdown.

To further improve the efficiency of implementing Winograd CONV in hardware, we transform Eq. \ref{equ:wino_tile} and express it in the form of General Matrix Multiplication (GEMM) shown in Eq. \ref{equ:wino_conv}. 
\vspace{-6pt}
\begin{equation}
\small
    Y_{k,\tilde{x},\tilde{y}}=A^T \left[ \sum_{c=1}^{C} U_{k,c} \odot V_{c,i,\tilde{x},\tilde{y}} \right] A
    \label{equ:wino_conv}
    \vspace{-6pt}
\end{equation}
where $U$ and $V$ represent $GgG^T$ and $B^TdB$, respectively; and the pair $(\tilde{x}, \tilde{y})$ indicates the coordinate of the input tile. Since we can split all the EWMMs in Eq. \ref{equ:wino_conv} into $(m+r-1) \times (m+r-1)$ independent GEMMs, both Winograd and Spatial CONV can be represented in the form of GEMM. With the uniform representation, we can instantiate one engine but support two CONV modes.

\vspace{-4pt}
\subsubsection{\textbf{Processing Engine (PE)}} 
We define $PI$, $PO$, and $PT$ as three dimensions of parallel factors in a PE. Figure \ref{fig:system_arch} shows that a PE contains a $PT \times PT$ array of GEMM cores, and each GEMM core is a $PI \times PO$ broadcast array. The parallel dimensions of Spatial and Winograd CONV are shown in Figure \ref{fig:system_arch}, where $PI$ and $PO$ can be scaled to meet the resources restrictions of different FPGA platforms. $PT$ is equal to $(m+r-1)$ which indicates the input tile size of Winograd CONV algorithm.


For each GEMM core, we unroll along input and output channel dimensions, which means we will broadcast $PI$ channels of the input feature maps and collect $PO$ channels of the output feature maps during the computation. In this fashion, one GEMM core is able to compute one General Matrix-Vector Multiplication (GEMV) in each clock cycle. By configuring the PE in Spatial mode, all the GEMM cores are merged into one large broadcast array; while in Winograd mode, each GEMM core is responsible for calculating one element of the EWMM operation shown in Eq. \ref{equ:wino_conv}.

\vspace{-4pt}
\subsubsection{\textbf{Load and Save Manager}} Given the reusable PE, the next step is to pass sufficient and organized data (DNN input feature maps and parameters) to perform efficient computations. A novel reconfigurable load and save manager is proposed to handle data supply for diverse data access and computation patterns of Winograd and Spatial CONV. 
For Spatial mode, the load manager directly loads input feature maps and broadcast them to the PE, while the save manager sums up the results from each row of GEMM cores and pass the partial sum to the accumulating buffer. 
For Winograd mode, the load manager performs an online input transform from input tile $d$ to $B^TdB$, and passes the transformed input to GEMM cores in the PE. The save manager also transforms the output tile with constant matrix $A$ and pass the transformed results to the accumulating buffer. Regarding DNN parameters for Winograd, we perform an offline transformation from pretrained DNN models.

\vspace{-4pt}
\subsubsection{\textbf{CONV Operation Partition}}
\label{subsubsec:dataflow}
Due to the limited memory resources on FPGA, feature maps and weights are not always accommodated by on-chip memory. To solve this problem, we use an adaptive partition strategy to ensure a flexible support for different FPGA platforms. We partition input and output feature maps into $H$ (Spatial CONV) or $H/m$ (Winograd CONV) groups along the dimension of feature map height $H$. We partition the DNN weights into $G_K$ groups along the dimension of output channels $K$. Under this strategy, one CONV operation is partitioned into $H \times G_K$ and $H/m \times G_K$ groups for Spatial and Winograd CONV, respectively. We also provide two types of dataflow for processing the CONV operation as IS and WS. For IS dataflow, the accelerator first loads one group of input feature maps, followed by $G_K$ groups of weights serially. It then completes all calculations related to the available input feature maps. For WS dataflow, accelerator keeps weights on chip, and for each group of weights, all groups of input feature maps need to be loaded to carry out the computation.

\vspace{-4pt}
\subsubsection{\textbf{Computing Mechanism}}
\begin{figure}[t]
    \centering
    \vspace{-20pt}
    \includegraphics[width=0.38\textwidth]{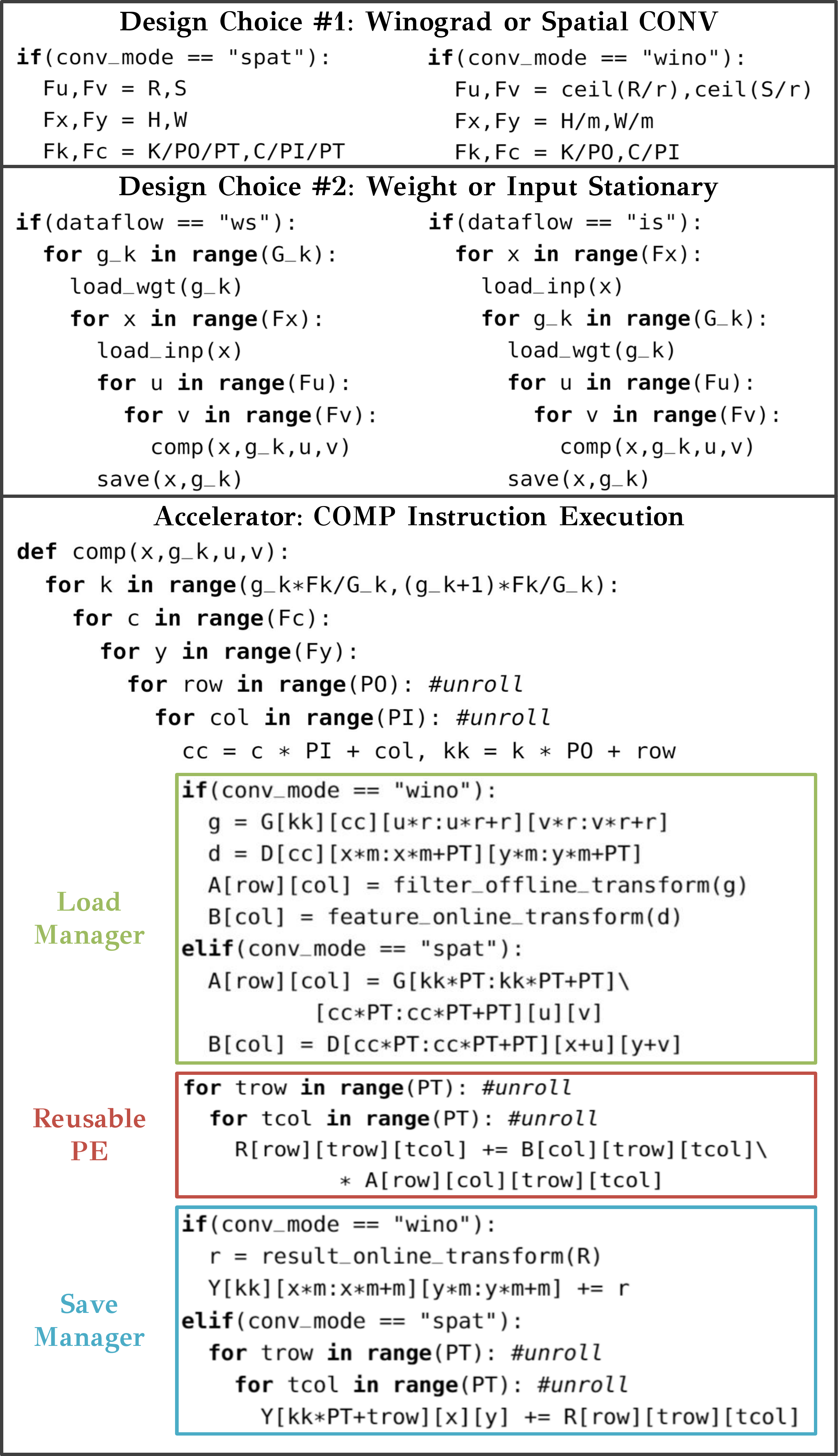}
    \vspace{-8pt}
    \caption{Pseudo-Code of the Execution of CONV}
    \vspace{-20pt}
    \label{fig:conv_pcode}
\end{figure}

Figure \ref{fig:conv_pcode} shows the pseudo-code of Spatial and Winograd CONV execution in our accelerator. Two design choices are first decided as: (1) CONV mode (Winograd or Spatial CONV) which determines the computing pattern of CONV operations; and (2) dataflow strategy (IS or WS) which determines the loops and instructions order and data dependency. 
The design choices here are often empirical which are mainly determined by computation/memory demands of DNN layers and available resources of the targeted FPGA. In general, the Winograd CONV requires higher memory access bandwidth than the Spatial one, while IS prefers larger feature maps compared to WS. In HybridDNN, we propose a novel DSE (Section \ref{sec:dse}) to leverage  these design choices and guarantee optimized performance.
With these information encoded in the instructions generated by HybridDNN compiler, our accelerator can flexibly execute CONV operations in four different design combinations but largely reuse the same PE without redundant structures. 


To enhance the generality of Winograd algorithm, we also introduce a kernel decomposition method in Figure \ref{fig:conv_pcode} to support larger kernel size ($> r \times r$) but using a $F(m \times m, r \times r)$ Winograd algorithm. Suppose we target a CONV layer with a $R \times S$ kernel ($R>r$, $S>r$), the kernel will be decomposed into $\left\lceil \frac{R}{r} \right\rceil \times \left\lceil \frac{S}{r} \right\rceil$ kernels with size $r \times r$, where zero padding will be applied if necessary. By accumulating the partial results of CONV with $r \times r$ kernels, we can output the same results of Wingrad CONV with larger kernel size. 

\vspace{-6pt}
\subsection{Memory Management}
\begin{figure}
    \centering
    \vspace{-20pt}
    \includegraphics[width=0.49\textwidth]{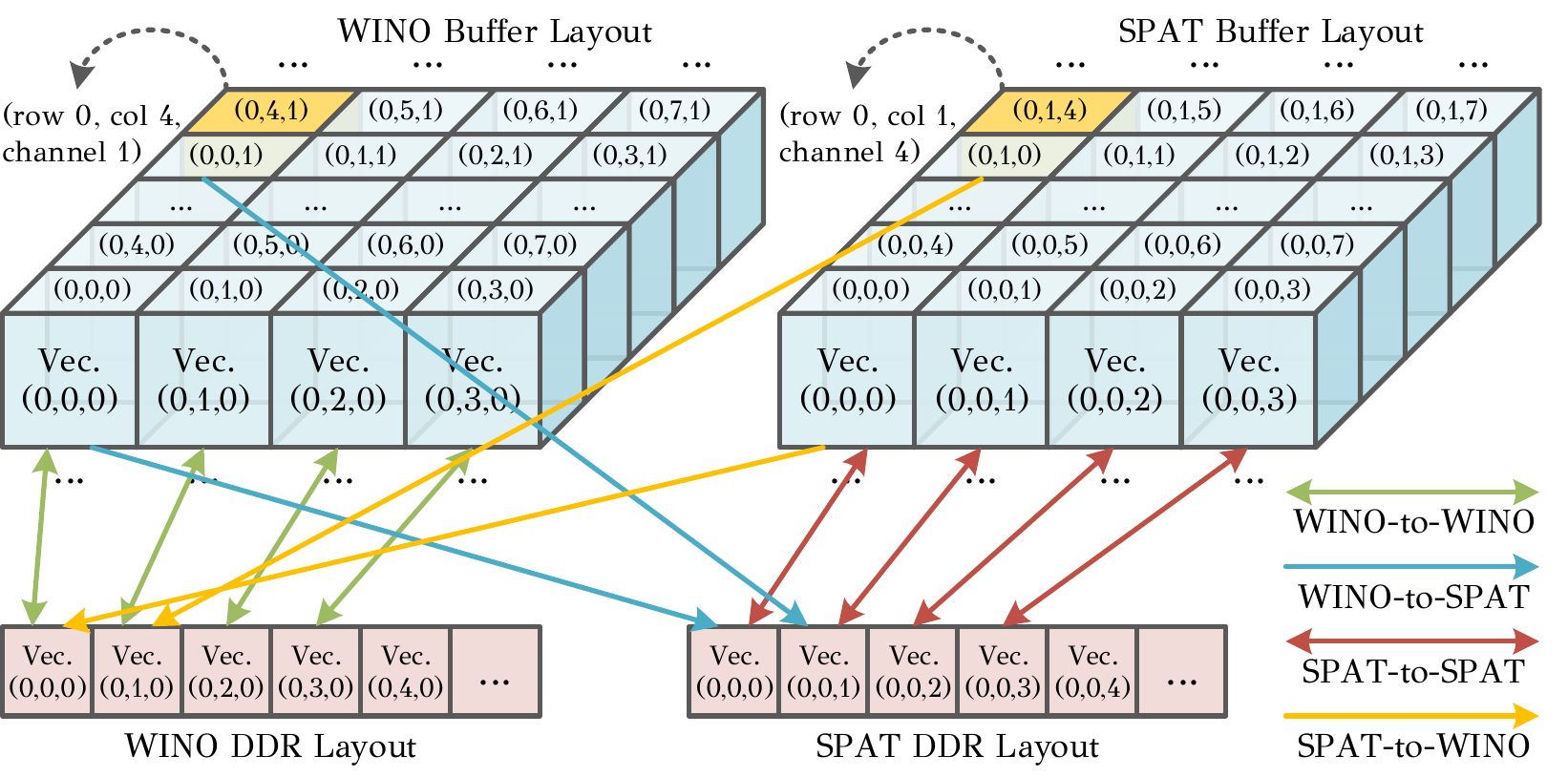}
    \vspace{-22pt}
    \caption{Feature Maps Data Layout}
    \label{fig:data_layout}
    \vspace{-10pt}
\end{figure}

\vspace{-2pt}
The memory access patterns are different for Spatial or Winograd CONV due to different parallel factors. It causes problems when successive layers are not implemented in the same CONV mode so that data reordering is inevitable between these layers.
To solve this problem, HybridDNN proposes a novel memory structure and an efficient data reordering mechanism.
The on-chip buffers are partitioned with factors shown in Table \ref{tab:buff_banks} to enable the parallel access to the data.
The data layouts in on-chip buffers and external memory are shown in Figure \ref{fig:data_layout} when $PT=4$.
We consider a GEMV operation of a vector with $PI$ channels of input feature maps as basic operation. So each element shown in Figure \ref{fig:data_layout} represents a vector of $PI$ or $PO$ channels for input or output feature maps, respectively.

Given this proposed data layout, we design a reconfigurable feature for the SAVE module to support all four possible data layout transforms (WINO-to-WINO, WINO-to-SPAT, SPAT-to-SPAT, and SPAT-to-WINO), while the LOAD module supports two transforms (WINO-to-WINO and SPAT-to-SPAT) as shown in Figure \ref{fig:data_layout}.
For example, we assume the first (left) and the second (right) layer are implemented by Winograd and Spatial CONV, respectively. For the first layer, the SAVE module will work at WINO-to-SPAT mode and pass the output feature maps to the external memory following the blue arrows. Then, for the second layer, the LOAD module will load the inputs from the external memory following red arrows. With this mechanism, the required data reordering is offloaded to the SAVE module, which ensures proper data layouts for different CONV modes chosen by the successive layer.

\begin{table}[t]
    \centering
    \caption{Partition Factors of On-Chip Buffers}
    \vspace{-10pt}
    \footnotesize
    \label{tab:buff_banks}
    \begin{tabular}{cccc}
        \toprule
         & \textbf{In Buffer} & \textbf{Weight Buffer} & \textbf{Out Buffer} \\
        \midrule
        \textbf{In Channel} & $PI(PI \times PT)$ & $PI(PI \times PT)$ & $-$\\
        \textbf{Out Channel} & $-$ & $PO(PO \times PT)$ & $PO(PO \times PT)$\\
        \textbf{Fmap Row} & $PT(1)$ & $-$ & $m(1)$ \\
        \textbf{Fmap Col.} & $PT(1)$ & $-$ & $m(1)$ \\
        \textbf{Weight Row} & $-$ & $PT(1)$ & $-$ \\
        \textbf{Weight Col.} & $-$ & $PT(1)$ & $-$ \\ 
        \bottomrule
        \multicolumn{4}{l}{\scriptsize \tabincell{l}{*Partition factors inside of the brackets belong to Spatial mode, while factors outside \\ of the brackets belong to Winograd mode.}}
    \end{tabular}
    \vspace{-15pt}
\end{table}


\vspace{-6pt}
\section{Performance Estimation and Design Space Exploration}
\label{sec:dse}
\vspace{-2pt}
With the scalable and flexible accelerator design paradigms, HybridDNN can explore a large design space in both hardware and software perspectives. In this section, we first develop highly-accuracy analytical models for resource utilization and latency estimation, and then design a DSE tool to effectively explore design space and find out the optimal configurations for building DNN accelerators.

\vspace{-6pt}
\subsection{Resources Utilization Modeling}
\vspace{-2pt}
The resource utilization of the generated accelerator is determined by $PI$, $PO$, $PT$, and $DATA\_WIDTH$.
Since HybridDNN supports $F(2 \times 2, 3 \times 3)$ and $F(4 \times 4, 3 \times 3)$ Winograd algorithms, $PT$ should be equal to 4 or 6. It is possible to support larger $PT$ in Winograd algorithm, but it also introduces a large amount of extra additions which eliminates the advantage of using Winograd mode \cite{zhuge}. The data bitwidth is represented by $DATA\_WIDTH$. 
Regarding DSP utilization, we use Eq. \ref{equ:dsp_util} by considering three factors as: (1) the size of PE; (2) the data bitwidth of the multiplier inputs; and (3) the number of multipliers for address generation. $\alpha$ is the correction term related to quantization strategies and $\beta$ is the number of DSPs for address generation, which is an FPGA-independent constant.
\vspace{-2pt}
\begin{equation}
    \label{equ:dsp_util}
    \small
    N_{DSP}=PI \times PO \times PT^2+ \alpha \times PO \times m^2+PO+ \beta
    \vspace{-4pt}
\end{equation}

For memory resource, we assume all the on-chip memory are implemented using BRAM. 
The utilization of BRAM mainly comes from on-chip buffers as shown in Eq. \ref{equ:bram_util}, where $BRAM\_WIDTH$ is the data bitwidth of one BRAM instance on the targeted FPGA.
\vspace{-2pt}
\begin{equation}
    \label{equ:bram_util}
    \small
    \begin{aligned}
        N_{BRAM}=\frac{DATA\_WIDTH}{BRAM\_WIDTH} \times \left (PI \times PT^2 \right. \\
        \left. +PI \times PO \times PT^2+ \left ( 1+\alpha \right ) \times PO\times m^2 \right )
    \end{aligned}
    \vspace{-4pt}
\end{equation}

We also estimate the LUT utilization as:
\vspace{-3pt}
\begin{equation}
    \label{equ:lut_util}
    \small
    N_{LUT}=\gamma \times \left (PI \times PO \times PT^2 \right )\times \left(1 + \delta \times m^2 \right)
    \vspace{-3pt}
\end{equation}
where $\gamma$ is the number of LUTs per MAC unit and $\delta$ is the correction term related to the impact of $m$.
In our design, $\alpha$, $\beta$, $\gamma$, and $\delta$ can be pre-defined through profiling.

\vspace{-6pt}
\subsection{Latency Modeling}
\vspace{-2pt}
The latency of executing a CONV layer using the proposed COMP module with working frequency $FREQ$ can be calculated by Eq. \ref{equ:T_comp_spat} (Spatial) and Eq. \ref{equ:T_comp_wino} (Winograd).
\vspace{-2pt}
\begin{equation}
    \label{equ:T_comp_spat}
    \small
    T_{CP}^{spat}=\frac{K \times C \times R \times S \times H \times W}{FREQ \times PI \times PO \times PT^2}
\end{equation}
\vspace{-2pt}
\begin{equation}
    \label{equ:T_comp_wino}
    \small
    T_{CP}^{wino}=\frac{K \times C \times \left\lceil \frac{R}{r} \right\rceil \times \left\lceil \frac{S}{r} \right\rceil \times PT^2 \times H \times W}{FREQ \times PI \times PO \times PT^2 \times m^2}
\end{equation}

The latency of using Winograd mode is lower than Spatial mode when running the same CONV layer due to lower computation complexity. Assuming the external memory bandwidth is $BW$, we use Eq. \ref{equ:T_load_wgt_spat} (Spatial) and Eq. \ref{equ:T_load_wgt_wino} (Winograd) to indicate the latency of loading data in LOAD\_WGT module.
\vspace{-2pt}
\begin{equation}
\small
    \label{equ:T_load_wgt_spat}
    T_{LDW}^{spat}=\frac{K \times C \times R \times S}{min(BW,FREQ \times PI \times PO \times PT)}
\end{equation}
\begin{equation}
    \label{equ:T_load_wgt_wino}
    \small
    T_{LDW}^{wino}=\frac{K \times C \times \left\lceil \frac{R}{r} \right\rceil \times \left\lceil \frac{S}{r} \right\rceil \times PT^2}{min(BW,FREQ \times PI \times PO \times PT)}
\end{equation}

Noted that Winograd mode asks more data from memory compared to Spatial mode, so the latency of loading data in Winograd mode is much longer. For example, assuming $m=4$ and $r=3$ with $5 \times 5$ kernel, the loading latency of Winograd mode is $\frac{2 \times 2 \times 6^2}{5 \times 5}=5.76\times$ compared to Spatial mode. We also calculate the latency of LOAD\_INP (Eq. \ref{equ:T_load_inp}) and SAVE module (Eq. \ref{equ:T_save}):
\begin{equation}
    \label{equ:T_load_inp}
    \small
    T_{LDI}=\frac{C \times H \times W}{min(BW,FREQ \times PI \times PT)}
\end{equation}
\begin{equation}
    \label{equ:T_save}
    \small
    T_{SV}=\frac{K \times H \times W}{min(BW,FREQ \times PO \times PT)}
\end{equation}

Assuming all the functional modules work concurrently, the overall latency is determined by the one with the longest latency. However, it is more complicated in reality because there exist data dependencies between these modules. Taking this and the computing mechanism described in subsection \ref{sec:engine} into consideration, the memory access latency that cannot be hidden is separately calculated as $T_{penalty}$.
So, the overall latency can be modeled as:
\vspace{-2pt}
\begin{equation}
    \label{equ:T_spat_is}
    \small
    T^{spat-is}=max \left( T_{LDI}, H \times T_{LDW}^{spat},T_{CP}^{spat},T_{SV} \right) +T_{penalty}^{spat-is}
\end{equation}
\begin{equation}
    \label{equ:T_spat_ws}
    \small
    T^{spat-ws}=max \left( G_K \times T_{LDI}, T_{LDW}^{spat},T_{CP}^{spat},T_{SV} \right) +T_{penalty}^{spat-ws}
\end{equation}
\begin{equation}
    \label{equ:T_wino_is}
    \small
    T^{wino-is}=max \left( T_{LDI}, \frac{H}{m} \times T_{LDW}^{wino},T_{CP}^{wino},T_{SV} \right) +T_{penalty}^{wino-is}
\end{equation}
\begin{equation}
    \label{equ:T_wino_ws}
    \small
    T^{wino-ws}=max \left( G_K \times T_{LDI}, T_{LDW}^{wino},T_{CP}^{wino},T_{SV} \right) +T_{penalty}^{wino-ws}
\end{equation}


\vspace{-10pt}
\subsection{Architecture-Aware DNN Mapping}
\vspace{-2pt}
With the accurate estimations of resource overhead and latency, the design space exploration becomes an optimization problem targeting the lowest overall latency of processing a specific DNN model. 
We use Table \ref{tab:opt_problem} to describe this optimization problem when targeting a DNN with $L$ CONV or FC layers. We assume the latency of the $l$-th layer is $T_l$. Also, we introduce a hardware parameter $NI$ to represent the number of accelerator instances on one FPGA.
%
To solve the optimization problem, we propose a novel DSE algorithm to search for the optimal design choice in 3 steps.
In \textbf{Step(1)}, given the limited choices of $PT$, for each $PT$, 
we take turns to increase the value of $PI$, $PO$, and $NI$ until any one of the resource constraints is no longer satisfied. We then collect the possible combinations regarding HW parameters listed in Table \ref{tab:opt_problem}. In \textbf{Step(2)}, we consider SW parameters (CONV modes and dataflows) on top of the possible combinations from the first step, and evaluate the layer latency using Eq. \ref{equ:T_spat_is}-\ref{equ:T_wino_ws}.
Finally, in \textbf{Step(3)}, we traverse all candidate choices from the second step and select the best one. 
Assuming Step(1) provides $N$ different hardware instance candidates,
the computation complexity of Step(2) and Step(3) should be $O(N \times L)$ and $O(N)$, respectively.

\begin{table}[]
    \centering
    \caption{DSE Optimization Problem}
    \vspace{-10pt}
    \footnotesize
    \label{tab:opt_problem}
    \begin{tabular}{cc}
        \toprule
        \textbf{HW Parameters} & $PI,PO,PT,NI$ \\
        \midrule
        \textbf{SW Parameters} & \tabincell{c}{$\{mode_1, mode_2, ... mode_L\},$\\$\{dataflow_1,dataflow_2, ... dataflow_L\}$} \\
        \midrule
        \textbf{Constraints} & \tabincell{c}{$PI \geq PO \geq 1,PT \in \{4,6\},$ \\ $N_{LUT}<LUT,N_{DSP}<DSP,N_{BRAM}<BRAM,$ \\ $mode_l \in \{"spat","wino"\},dataflow_l \in \{"is","ws"\}$ } \\
        \midrule
        \textbf{Objective} & $\sum_{l=1}^{L}T_l$ \\
        \bottomrule
    \end{tabular}
    \vspace{-4pt}
\end{table}

\vspace{-6pt}
\section{Experimental Results}
\vspace{-2pt}
For demonstration, HybridDNN targets a cloud FPGA (Semptian NSA.241 with Xilinx VU9P) and an embedded FPGA (Xilinx PYNQ-Z1) for generating DNN accelerators. 
%
%
%
For the cloud design, 
input images are sent through PCIe to the on-board DDR4, while the output results are collected by the host CPU.
%
For the embedded design using PYNQ-Z1, the inputs are copied from SD card to main memory and processed by the programmable logic (PL). Results are then sent back to the processor system (PS).

\vspace{-6pt}
\subsection{VGG16 Case Study}
\vspace{-2pt}
We implement accelerators for running VGG16 on VU9P and PYNQ-Z1.
Since VU9P has three dies (which shares the multi-dies feature in latest cloud FPGAs), HybridDNN generates six accelerator instances (with configuration: $PI=4$, $PO=4$, and $PT=6$) to match the number of dies as two instances can fit in one die.
For the design on PYNQ-Z1, HybridDNN generates one accelerator instance with the configuration of $PI=4$, $PO=4$, and $PT=4$. 
Since using Winograd algorithm can be beneficial to process the CONV layers in VGG16, the proposed DSE select Winograd CONV for both designs, and we present the resource utilization in Table \ref{tab:hw_config}.
Compared to the conventional architecture which only supports Spatial CONV, the overhead of adding Winograd supported hybrid structure (including the Winograd transformation and the reconfiguration features of the functional modules) costs only 26.4\% extra LUTs but no extra DSPs on a VU9P FPGA. The main reason is that, in HybridDNN, Spatial and Winograd CONV can reuse the same PE to avoid  wasting resources. This feature also helps to exploit available FPGA resources better, as most of the FPGA-based DNN accelerators mainly rely on DSP resources while leaving a considerable amount of LUTs unused.

\begin{table}[t!]
    \centering
    \vspace{-5pt}
    \caption{Resource Utilization of VU9P and PYNQ-Z1}
    \label{tab:hw_config}
    \vspace{-10pt}
    \footnotesize
    \begin{tabular}{ccccc}
        \toprule
         & \textbf{LUTs} & \textbf{DSPs} & \textbf{18Kb BRAMs} \\
        \midrule
        \textbf{VU9P} & 706353 (59.8\%) & 5163 (75.5\%) & 3169 (73.4\%) \\
        \textbf{PYNQ-Z1} & 37034 (69.61\%) & 220 (100\%) & 277 (98.93\%) \\
        \bottomrule
    \end{tabular}
    \vspace{-15pt}
\end{table}

\vspace{-6pt}
\subsection{Evaluation and Comparison}
\begin{figure}[t]
    \centering
    \vspace{-15pt}
    \includegraphics[width=0.42\textwidth]{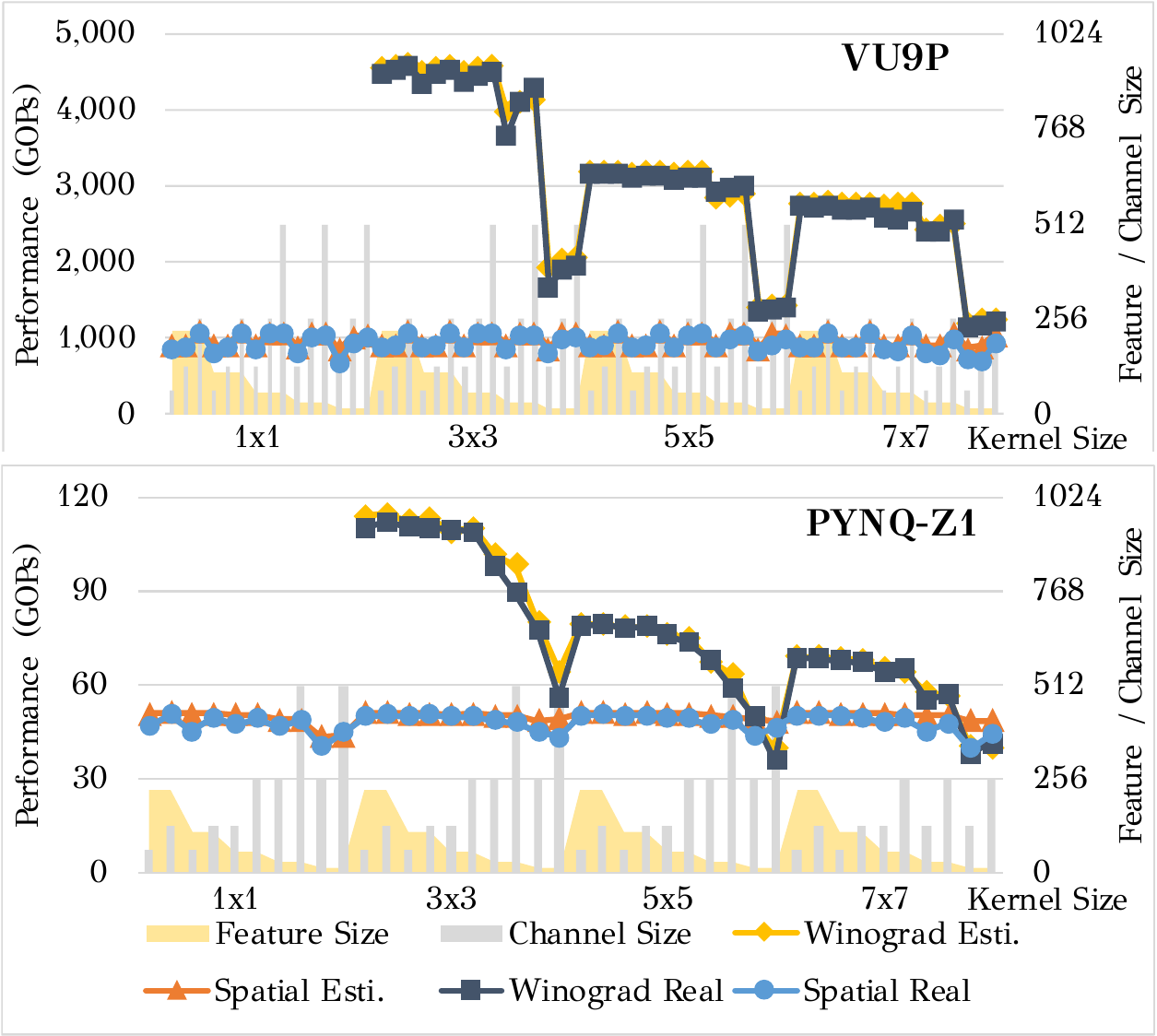}
    \vspace{-10pt}
    \caption{Performance of VU9P and PYNQ-Z1}
    \label{fig:test_perf}
    \vspace{-15pt}
\end{figure}

\vspace{-2pt}
In Table \ref{tab:vgg_result}, we compare the performance of HybridDNN to previously published works on VGG16 model. Results show that our design for VU9P achieves a 1.8x higher performance (GOPS) and 2.0x higher energy efficiency compared to the state-of-art DNN accelerator implemented using the same FPGA.
To evaluate the flexibility of HybridDNN, we extend our test cases and evaluate 60 and 40 CONV layers
(with different feature map size, channel number, and kernel size) using the generated accelerators targeting VU9P and PYNQ-Z1, respectively. Results in Figure \ref{fig:test_perf} indicate the performance of Spatial mode is stable and close to the peak achievable performance, while the performance of Winograd mode presents certain patterns that fluctuates across different CONV layers. 
By handling the same CONV layer, Winograd mode spends less computation time than Spatial mode, which equivalently causes higher demands of memory access bandwidth (as the same amount of DNN parameters needs to be loaded from DRAM within a smaller time-slot). When a memory-bound is encountered, the performance of Winograd mode drops. In the VGG16 case study, the DSE selects all CONV layers of VGG16 to be implemented in Winograd mode due to the sufficient memory bandwidth. However, in other scenarios (e.g., IoT applications) where the available memory bandwidth is limited by the embedded devices, Spatial CONV may outperform Winograd. The flexible support for both Spatial and Winograd CONV allows the proposed HybridDNN framework to deliver the optimal solutions and fit into a wide range of different scenarios.
We also compare the estimated results from our proposed analytical models to the HybridDNN generated hardware implementation results, and only 4.27\% and 4.03\% errors are found for accelerators running on VU9P and PYNQ-Z1, respectively. The accurate estimations guarantee valid design space explorations in HybridDNN.

\vspace{-6pt}
\section{Conclusions}
\vspace{-2pt}
In this paper, we presented HybridDNN, a framework for building DNN accelerators on FPGAs with high-performance and energy efficiency. We proposed a highly scalable DNN accelerator architecture for efficient deployment onto cloud/embedded FPGAs and a flexible PE structure with hybrid Spatial/Winograd CONV support for diverse CONV layers. We designed a comprehensive analytical tool for fast and accurate performance estimation (with 4.27\% and 4.03\% error rate for accelerator designs in VU9P and PYNQ-Z1, respectively), and a DSE engine to provide the best configurations regarding CONV modes (Spatial/Winograd), dataflows (IS/WS), and parallel factors. With the above novel technologies, HybridDNN delivered accelerators with the highest performance peaking at 3375.7 (VU9P) and 83.3 (PYNQ-Z1) GOPS, and the best energy efficiency (73.5 GOPS/W) compared to previously published results. 

\begin{table}[t]
    \centering
    \vspace{-15pt}
    \caption{Comparison with Previous Works}
    \label{tab:vgg_result}
    \vspace{-10pt}
    \footnotesize
    \begin{tabular}{c|ccc|cc}
        \toprule
         & \textbf{\cite{tgpa}} & \textbf{\cite{improving}} & \textbf{\cite{clouddnn}} & \multicolumn{2}{c}{\textbf{Ours}} \\
        \hline
        \textbf{Device} & \tabincell{c}{Xilinx\\VU9P} & \tabincell{c}{Arria10\\GX1150} & \tabincell{c}{Xilinx\\VU9P} & \tabincell{c}{Xilinx\\VU9P} & \tabincell{c}{PYNQ\\Z1} \\
        \hline
        \textbf{Model} & VGG16 & VGG16 & VGG16 & VGG16 & VGG16 \\
        \hline
        \textbf{Precision} & 16-bit & 16-bit & 16-bit & 12-bit* & 12-bit* \\
        \hline
        \textbf{Freq.(MHz)} & 210 & 385 & 214 & 167 & 100 \\
        \hline
        \textbf{DSPs} & 4096 & 2756 & 5349 & 5163 & 220 \\
        \hline
        \textbf{\tabincell{c}{CNN\\Perf.(GOPS)}} & 1510 & 1790 & 1828.6 & \textbf{3375.7} & 83.3 \\
        \hline
        \textbf{Power(W)} & NA & 37.5 & 49.3 & 45.9 & 2.6 \\
        \hline
        \textbf{\tabincell{c}{DSP Effi.\\(GOPS/DSP)}} & 0.37 & 0.65 & 0.34 & \textbf{0.65} & 0.38 \\
        \hline
        \textbf{\tabincell{c}{Energy Effi.\\(GOPS/W)}} & NA & 47.78 & 37.1 & \textbf{73.5} & 32.0 \\
        \bottomrule
        \multicolumn{6}{l}{\scriptsize \tabincell{l}{*DNN parameters are quantized to 8-bit; input feature maps are set to 12-bit \\in PE due to the Winograd matrix transformation}}
    \end{tabular}
    \vspace{-10pt}
\end{table}

\vspace{-5pt}
\begin{acks}
\vspace{-2pt}
This work is supported in part by the IBM-Illinois Center for Cognitive Computing Systems Research (C3SR) and XMotors.ai.
\vspace{-3pt}
\end{acks}

\bibliographystyle{unsrt}
\footnotesize
\bibliography{literature}

\end{document}